\documentclass[twocolumn,amssymb,amsmath,showpacs,aps,pre]{revtex4-1}
\usepackage[dvips]{graphicx}
\usepackage{epsfig}

\begin{document}

\title{Intermittent pathways towards a dynamical target}

\author{F{\'e}lix Rojo}

\thanks{Fellow of Consejo Nacional de
Investigaciones Cient{\'{\i}}ficas y T{\'e}cnicas}
\author{Pedro A. Pury}
\author{Carlos E. Budde}

\affiliation{Fa.M.A.F., Universidad Nacional de C\'ordoba,\\
Ciudad Universitaria, X5000HUA C\'ordoba, Argentina}

\date{Received: 13 October 2010}

\begin{abstract}
In this paper, we investigate the quest for a single target,
that remains fixed in a lattice, by a set of independent walkers.
The target exhibits a fluctuating behavior between trap and
ordinary site of the lattice, whereas the walkers perform
an intermittent kind of search strategy.
Our searchers carry out their movements in one of two states
between which they switch randomly.
One of these states (the exploratory phase) is a symmetric nearest
neighbor random walk and the other state (relocating phase) is a
symmetric next-nearest neighbor random walk. By using the multistate
continuous-time random-walk approach we are able to show that
for dynamical targets, the intermittent strategy (despite the
simplicity of the kinetics chosen for searching) improves detection,
in comparison to displacements in a single state.
We have obtained analytic results, that can be numerically evaluated,
for the Survival Probability and for the Lifetime of the target.
Thus, we have studied the dependence of these quantities both
in terms of the transition probability that describes the
dynamics of the target and in terms of the parameter that
characterizes the walkers' intermittency.
In addition to our analytical approach, we have implemented
Monte Carlo simulations, finding excellent agreement between
the theoretical--numerical results and simulations.

\end{abstract}


\maketitle

%%%%%%%%%%%%%%%%%%%%%%%%%%%%%%%%%%%%%%%%%%%%%%%%%%%%%%%%%%%%%%%%%%%%%%
\section{Introduction}
%%%%%%%%%%%%%%%%%%%%%%%%%%%%%%%%%%%%%%%%%%%%%%%%%%%%%%%%%%%%%%%%%%%%%%

Search problems have recently experienced a rapid growth and
motivated a great deal of work in the most various situations
(see Ref.~\cite{000} and references therein): fishermen and shoals,
prey and predators, two molecules in the course of a reaction,
a protein that looks for an specific site on DNA strand,
medical drugs and illnesses.
Thus, search problems spans over a wide range of domains and
fields. When the target if fixed at a given location in space,
the search problem is equivalent to the trapping problem, i.e.,
the situation where a set of walkers independently diffuse in
space until one of them is caught by the trap.

Among different forms of search strategies~\cite{000,010,*012,020},
the so called {\em intermittent strategies}, which combine a phase
of relocation (where the searcher may or may not be capable to
capture the target), with a phase of search (where target capture
is always allowed), have been proved relevant and able to be
optimized at various scales. Intermittent motion occurs in a wide
array of living organisms from protozoans to mammals. It has been
observed that numerous animal species switch between two distinct
types of behavior while foraging or searching for shelter,
or mate \cite{025,026,027}.
At a microscopic scale, we find intermittent motion, e.g.,
in the binding of a protein to specific sites on DNA for regulating
transcription, as it is the case when the protein has the ability
of diffuse in one dimension by sliding along the length of the DNA,
in addition to their diffusion in bulk
solution~\cite{030,035,040,045}.

In Refs.~\cite{050,055,060} a theoretical model for the
search kinetics of a hidden target was presented, assuming that each
searcher could be in either of two states of diffusion.
It was shown that intermittent strategies always improve target
detection in comparison with the single--state displacement.
An important aspect in the searching dynamics that has recently
been studied~\cite{055,070} is to consider not only the different
states in the motion of walkers, but also what happens with the
trapping process when additional dynamical effects are taken
into account. For instance, in Ref.~\cite{070,055} the sighting
range and the smell capacity of predators was considered as a sort
of additional search ability.

The aim of this study is to complete and to extend previous
results~\cite{050,055,060} and to present another relevant case
for modeling real search problems: The inclusion of a fluctuating
behavior in the target. These fluctuations can modify and prevent
the encounter between a searcher and the target to be successful.
This behavior may be due to the internal evolution of the trap
or due to their interaction with a changing environment.
For instance, in a chemical context, the activation or deactivation
of a reagent can be caused by external factors (photons,
solvent molecules, etc.)~\cite{080}.
In biological contexts, the dynamical behavior of the target
is also a determinant, e.g., reactions occurring within biomembranes
require some geometrical configurations in the biomolecule structure
to be completed. The absent of these configurations inhibit the
reaction, whereas stochastic changes in the molecule geometry
can let it take place. Even the delivery of drug in medical
treatments can involve blocking chemical reactions, in order
to boost the delivered medicine effectiveness~\cite{090}.

Ref.~\cite{100} introduced a generalization of the trapping model
which allows encounters between particles of two kinds,
$A$ and $B$, with or without annihilation depending on the internal
state of the particle $A$.
Particle $A$, which is identified as the trap in this work,
has two states: An active one in which the annihilation does
take place, and an inactive one in which it behaves as a regular
site. The particles $B$ are our walkers.
In this work, we take a step forward in modeling search problems by
the formulation of an unified framework which comprises the dynamical
behavior of the trap and the intermittent search strategy performed
by the walkers.
We exploit the theory of \textit{multi--state} random--walk
(RW)~\cite{110}, we use the concepts of \textit{Survival
Probability} (SP) and \textit{Mean Target Lifetime} (MTL),
and we establish the connection to the \textit{First--Passage Time}
(FPT) corresponding to the problem of one walker.

The outline of this paper is as follows. The next section presents
our model and gives the basic definitions and concepts.
Also, this Section describes the analytical approach to the trapping
process.
Section III presents the main results for the SP and the MTL
of the target through a comparison between the numerical evaluation
of our analytical framework and Monte Carlo simulations. In Sec.~IV
we discuss our conclusions. Finally, in Appendix A we develop
the analytical calculations of Sec.~III, corresponding to infinite
chains and rings, whereas in Appendix B we consider the high
transition regime for the trap.

%%%%%%%%%%%%%%%%%%%%%%%%%%%%%%%%%%%%%%%%%%%%%%%%%%%%%%%%%%%%%%%%%%%%%%
\section{Analytical Approach}
%%%%%%%%%%%%%%%%%%%%%%%%%%%%%%%%%%%%%%%%%%%%%%%%%%%%%%%%%%%%%%%%%%%%%%

\subsection{The Model}

We restrict our work to chains (finite and infinite) and assume
that the dynamical trap is held fixed at the origin of the lattice.
A set of walkers, uniformly distributed along the chain,
starts the ``search'' at $t=0$.
At the trap site, the following situations may occur:
\begin{itemize}
\item The trap is in an active status, i.e., it works
(the first walker reaching the trap is caught with probability
one, i.e.,perfect trapping).

\item The trap is in a passive status and stays that way
until the searcher leaves it, i.e., the trap behaves like any other
chain site, and capture can not be carried out.

\item The trap is in a passive status but changes its
condition before the searcher leaves, i.e., the capture
is also performed.
\end{itemize}
We denote the internal states of the trap by $i=1$ (active status)
and $i=2$ (passive status).
On the other hand, we assume that each predator can make two types
of motion on the lattice:
\begin{itemize}
\item {\em Exploration:} RW with symmetric jumps to first
nearest neighbors, with transition probability per unit time
$\lambda$, and

\item {\em Relocalization:} RW with symmetric jumps to second
nearest neighbors, also with transition probability per unit time
$\lambda$.
\end{itemize}
We also assume that the walkers' dynamics and the dynamical behavior
of the trap are independent.

\begin{figure}[!hpt]
\begin{center}
% two columns
\includegraphics[clip,width=0.45\textwidth]{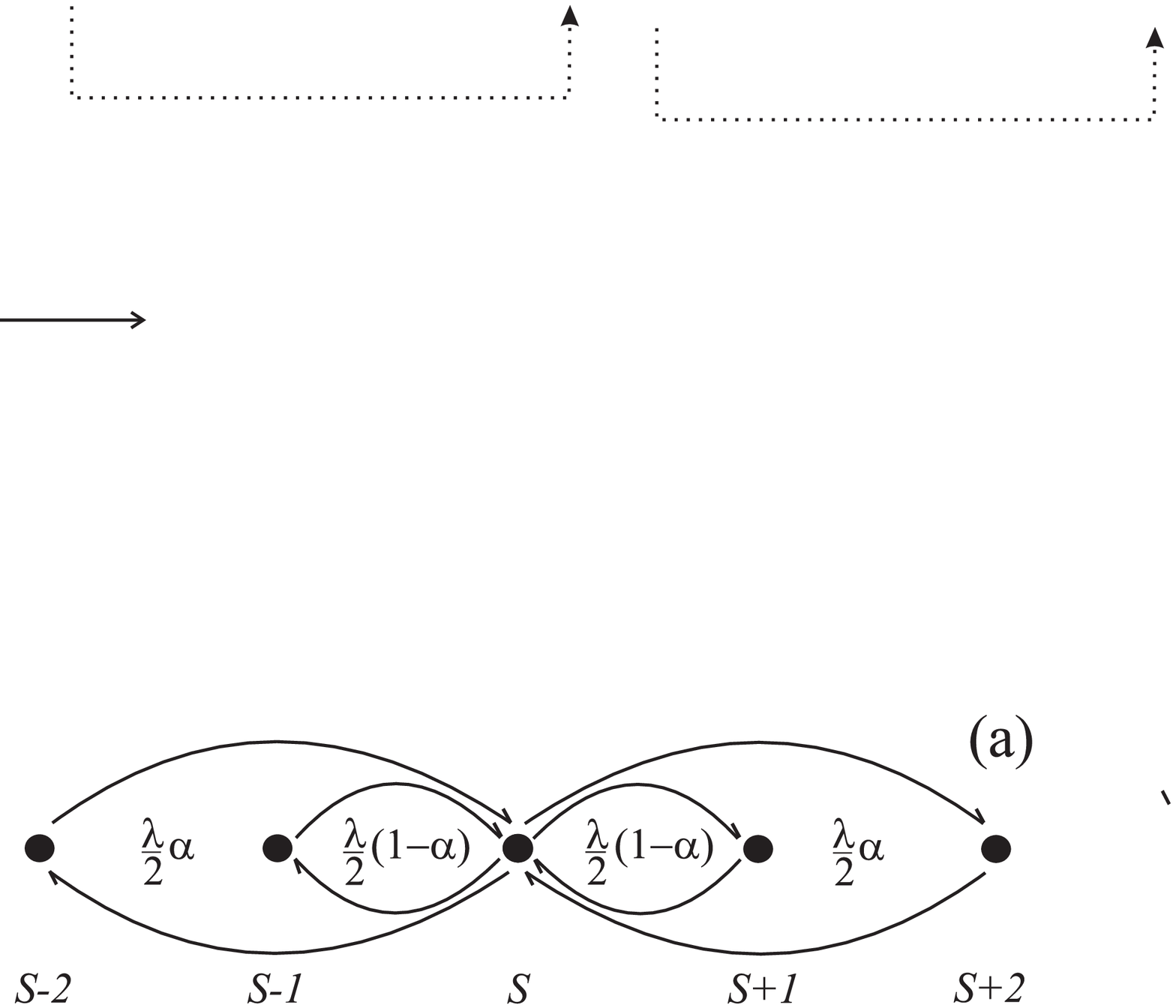}\vspace{0.5cm}
\includegraphics[clip,width=0.45\textwidth]{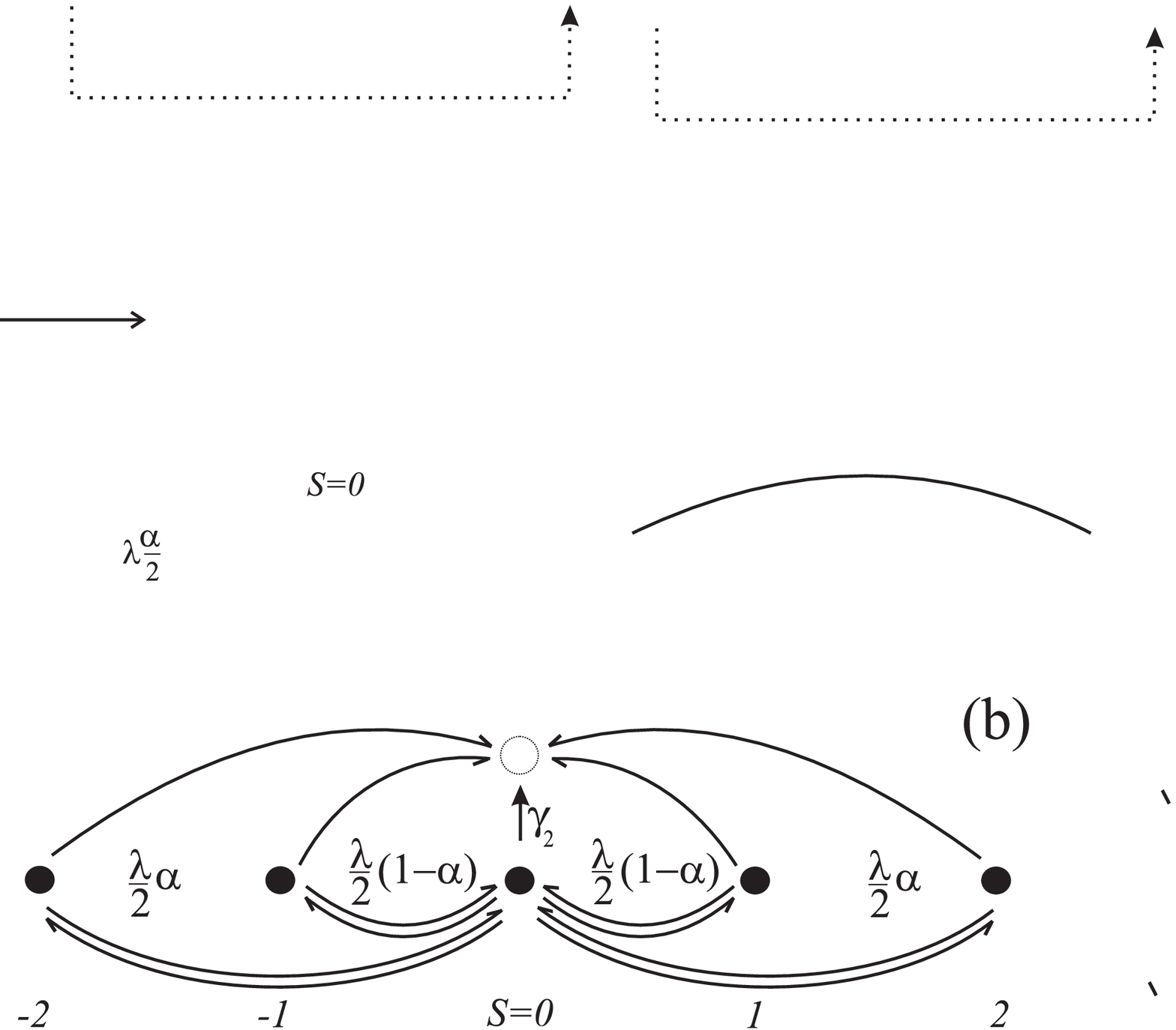}
% one column
%\includegraphics[clip,width=0.90\textwidth]{Fig_1a.eps}\vspace{0.5cm}
%\includegraphics[clip,width=0.90\textwidth]{Fig_1b.eps}
%
\caption{\textbf{(a)} Schematic transitions of the walker
to/from site $s$ (away from the trap: $s \neq -1, 0, 1$) and
\textbf{(b)} Walker transitions to/from $s=0$ (trap site).
A walker dwelling at site $0$ could be trapped with rate
$\gamma_2$ (the probability transition rate for activation of
the trap). The dynamics of the trap is independent of the
dynamics of walkers.}
\label{Fig1}
\end{center}
\end{figure}

The proposed composite process can be described by the coupled master
equations
\begin{eqnarray}\label{mastereq1}
% \nonumber to remove numbering (before each equation)
\frac{\partial P_{1,i_0}(\vec{s},t|\vec{s}_0,0)}{\partial t}
&=& \mathbb{A}P_{1,i_0}(\vec{s},t|\vec{s}_0,0)
+ \gamma_2  \,P_{2,i_0}(\vec{s},t|\vec{s}_0,0)
\nonumber \\
&& -\gamma_1 \,P_{1,i_0}(\vec{s},t|\vec{s}_0,0) \,, \\
\label{mastereq2}
\frac{\partial P_{2,i_0}(\vec{s},t|\vec{s}_0,0)}{\partial t}
&=& \mathbb{A}P_{2,i_0}(\vec{s},t|\vec{s}_0,0)
+ \gamma_1  \,P_{1,i_0}(\vec{s},t|\vec{s}_0,0)
\nonumber \\
&&-\gamma_2 \,P_{2,i_0}(\vec{s},t|\vec{s}_0,0) \,,
\end{eqnarray}
where $P_{i,i_0}(\vec{s},t|\vec{s}_0,t=0)$ is the the conditional
probability of the walker of being at site $\vec{s}$ with the trap
in state $i$ at time $t$,  given that it was at site $\vec{s}_0$
with the trap in state $i_0$ at $t=0$.
For simplicity, we have restricted the {\em activation - deactivation}
process  of the trap to time exponential density functions with
parameters $\gamma _i$, i.e., $\gamma _i$ is the probability
transition rate of the trap to make a transition from its state $i$
to the other state.
The dynamical evolution of the walkers, taking into account its
intermittency, is described by the operator $\mathbb{A}$.
Particularly, for a chain, we get
\begin{eqnarray}\label{Aoperator}
[\mathbb{A}]_{s,s'} &=& \frac{\lambda}{2}
\left[ (1-\alpha)(\delta_{s,s'-1} + \delta_{s,s'+1})
\right.
\nonumber \\
&& \left.
+ \alpha(\delta_{s,s'-2} + \delta_{s,s'+2})
- 2 \,\delta_{s,s'} \right] \,,
\end{eqnarray}
where $\alpha$ is the parameter that regulates the walker's
frequency intermittency and $\lambda$ its diffusion constant
(see Fig.~\ref{Fig1}).

\subsection{The Trapping Process}\label{TrappingProcess}

We will focus on the SP of the dynamical target, i.e., the probability
that the target remains undetected up to a time $t$, and its closely
related quantity, the MTL~\cite{120}, which compute the time in which
the first walker reaches the target under the appropriate
circumstances of capture.
We define $F_{1,i_0}(\vec{0}, t|\vec{s}_0,0)$  as the First
Passage time density through the site $\vec{0}$ at time $t$, when
capture is possible, given that the searcher was at $\vec{s}_0$
with the target in state $i_0$ at time $t=0$. In the way of
Ref.~\cite{090}, we introduce  the notion of \textit{generalized
state} which takes into account the position of the walker and
the state of the target, $(\vec{s},i)$.
The connection between FPT density at $(\vec{0},1)$ at time $t$
from $(\vec{s}_0,i_0)$, $F_{1,i_0}(\vec{0}, t|\vec{s}_0,0)$,
and the conditional probability $P_{i,i_0}(\vec{s},t|\vec{s}_0,t=0)$
is established by
\begin{equation}\label{F1P}
\hat{F}_{1,i_0}(\vec{0},u|\vec{s}_0,0)=
\frac{\hat{P}_{1,i_0}(\vec{0},u|\vec{s}_0,0)}
{\hat{P}_{1,1}(\vec{0},u|\vec{0},0)} \,,
\end{equation}
which is the known Siegert's formula~\cite{130}, generalized
to internal states~\cite{140}.
We are denoting the Laplace transform of a function
of $t$ by a caret over the corresponding function. Thus, for example,
\begin{eqnarray}
\hat{P}_{i,i_0}(\vec{s},u|\vec{s}_0,0) &=&
\mathcal{L}\{P_{i,i_0}(\vec{s},t|\vec{s}_0,0)\}
\nonumber \\
&=& \int_0^{\infty}e^{-ut}P_{i,i_0}(\vec{s},t|\vec{s}_0,0) \,dt \,.
\nonumber
\end{eqnarray}

When trapping occurs independently of the \emph{initial} state
of the target, the SP in presence of only one walker may be written
(if $\vec{s}_0\neq\vec{0}$) as
\begin{equation}\label{PhiII}
\Phi_{1}(\vec{0},t|\vec{s}_0,0) = 1 - \sum_{i_0=1}^2 \theta_{i_0}
\int_{0}^{t} F_{1,i_0}(\vec{0},\tau|\vec{s}_0,0) \,d\tau \,.
\end{equation}
$\theta_{i_0}$ is the probability of the initial state of the target.
Thus, the target is initially active with probability $\theta_{1}$
or inactive with probability $\theta_{2}$.
The SP at time t, $\Phi_{N}(t)$, of the dynamical target
at the origin in the presence of $N$ independent walkers that
diffuse on an $M$-sites lattice can be written in terms of the SP
in the presence of only one walker, $\Phi_{1}(\vec{0},t|\vec{s}_0,0)$
as~\cite{120}
\begin{equation}\label{SurvivalN}
\Phi_N(t) =
\left( 1-\frac{1}{M-1}\sum_{\vec{s}_0\neq\vec{0}}
(1-\Phi_1(\vec{0}, t|\vec{s}_0,0)) \right)^N \,,
\end{equation}
where we have assumed a uniform probability distribution
for the initial position of the walkers, i.e, the probability that
a given walker is initially at a particular site
$\vec{s}_0(\neq\vec{0})$ is $(M-1)^{-1}$. Notice that we explicitly
exclude the possibility of having a walker at the position of the
target at $t=0$.
In the {\em bulk limit}, $N \rightarrow \infty$,
$M \rightarrow \infty$,
with $N/M \rightarrow \rho$, where the constant $\rho$
is the concentration of walkers, we get
\begin{equation}
\Phi_{\rho}(t) = \exp \left( -\rho \sum_{\vec{s}_0\neq\vec{0}}
( 1 - \Phi_1(\vec{0}, t|\vec{s}_0,0)) \right) \,.
\label{Phi_rho}
\end{equation}
The MTL is defined in finite domains as~\cite{120}
\begin{equation}\label{MTLN}
T_N = \int_0^{\infty}\Phi_N(t)dt \,,
\end{equation}
and in the {\em bulk limit} as
\begin{equation}\label{MTLrho}
T_{\rho} = \int_0^{\infty}\Phi_{\rho}(t)dt \,.
\end{equation}

We left for the Appendixes the detailed calculations of the
magnitudes presented in this section for the cases of infinite
chains and rings of $M$ sites.

%%%%%%%%%%%%%%%%%%%%%%%%%%%%%%%%%%%%%%%%%%%%%%%%%%%%%%%%%%%%%%%%%%%%%%
\section{Results}
\label{results}
%%%%%%%%%%%%%%%%%%%%%%%%%%%%%%%%%%%%%%%%%%%%%%%%%%%%%%%%%%%%%%%%%%%%%%

In this section, we illustrate the general framework presented
in the previous section. We consider one-dimensional systems
and give some general ideas to interpret the obtained results.
The inverse Laplace transform involved in the analytical
expressions, given in the Appendixes, is evaluated
numerically~\cite{145} for obtaining concrete results and then
we establish a comparison with independent Monte Carlo simulations.

A brief review of our simulation methodology is appropriate at this
point. We uniformly distribute the searchers (with probability
$\rho$ per site) in a one-dimensional lattice with periodic boundary
conditions.
The target is placed at the origin of the lattice. The propagation
of the searchers in the presence of a dynamical target is implemented
as follows. Each searcher has assigned an internal clock (all start
synchronized at time $t = 0$) which is updated according to their
waiting time probability distributions. For the {\em activation -
deactivation} process of the target a similar procedure to the
searchers is used; the target has assigned his own internal
clock which is updated with time exponential density functions
with parameters $\gamma _1$ and $\gamma _2$. We define an indicator
function that records the needed information: if the target was
captured up to a certain time (for the SP) and whether the target
was captured and the time in which this happened (for the mean target
lifetime). A randomly chosen walker take a step, to its nearest
neighbors with probability $(1-\alpha)$ or next-nearest neighbors
with probability $\alpha$ and left or right with equal probability
$(1/2)$. We check if the trapping conditions are fulfilled, and
if it does, we stop the dynamics, update our indicator function,
and generate a new ensemble of walkers. If it does not, we continue
the dynamics by taking another randomly chosen walker. Again,
if trapping occurs, the indicator function is updated and the
dynamics stopped; if not, the walk continues.
The output of interest of each realization is, for the SP, whether
it was captured up to a certain (predefined) time, and the time of
capture for the mean target lifetime.

In this section, we show the numerical results obtained
from the analytical expressions for the infinite chain
(see Appendix~\ref{InfiniteApp}) and the finite ring
(see Appendix~\ref{RingApp}).
For both the infinite and finite cases a
searchers' concentration $\rho=0.1$ was used
(for the finite case the concentration is defined by $\rho=N/M$).
All the finite chains considered in the figures correspond
to a ring of $M=20$ sites, with exception of Fig.~\ref{FigT2},
where different sizes ($M$) of the ring are explicitly stated.
All times are given in units of the inverse of
the diffusion constant ($1 /\lambda$).
It is worth to comment that when we talk about target transition
rates $\gamma_i$, these are \emph{low (high)} relative to the
diffusion constant $\lambda$.
An equivalent interpretation can be made if we consider the
target mean sojourn time in state $i$ as $\gamma_i^{-1}$;
this will be \emph{long (short)} on the time scale determined
by the propagator $\mathbb{A}$ ($\lambda^{-1}$).
In the following we will consider symmetric transition rates
for the {\em activation--deactivation} process of the target,
$\gamma_1=\gamma_2=\gamma$.

\begin{figure}[!tbp]
\begin{center}
% two columns
\includegraphics[clip,width=0.45\textwidth]{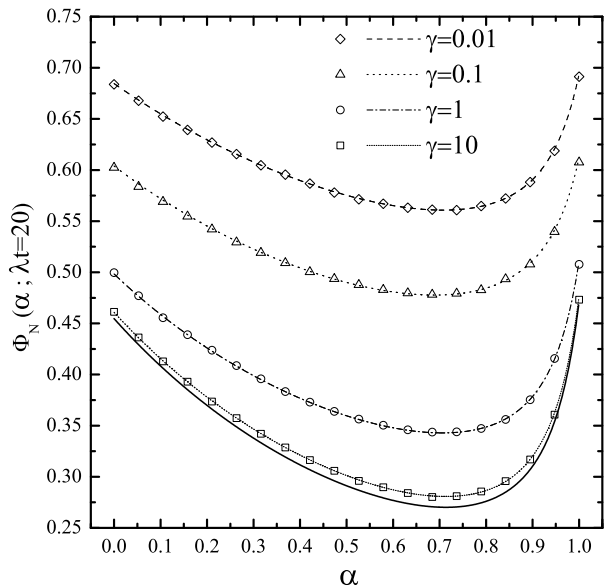}
% one column
%\includegraphics[clip,width=0.90\textwidth]{Fig_2.eps}
%
\caption{Analytical-Numerical calculations (lines) and Monte Carlo
simulations (symbols) for the SP, $\Phi_N(\alpha;t)$,
up to time $t=20$ for different target transition rates $(\gamma)$.
(I) (diamonds) $\gamma=0.01$, (II) (triangles) $\gamma=0.1$, (III)
(circles) $\gamma=1$ and (IV) (squares) $\gamma=10$. We have also
included for comparison the static trap case (thick solid line).}
\label{FigS1}
\end{center}
\end{figure}

In Fig.~\ref{FigS1} we present curves (for the finite case)
corresponding to  $\Phi_N(\alpha,t)$ for a fixed evolution
time $t=20$. Notice how the intermittent search can improve
the detection probability, i.e, minimize the SP of the target,
compared with the \textit{single state search} ($\alpha\sim 0$,
$\alpha\sim 1$).
As a comparison we also include the ``static trap'' case, i.e.,
the target is always active.
As can be seen from the figure, an optimal value for $\alpha$ can be
found for each target transition rates $\gamma$ chosen. Even though
all curves present a similar behavior, it is apparent that the
transition rate $\gamma$ plays an important role.
The ratio between the maximum value of the SP (at $\alpha=1$)
and it's minimum is almost of $80\%$ ($120 \%$) for $\gamma= 0.1$
($\gamma=0.01$). At high values of $\gamma$, the ``static trap''
case is approached.

\begin{figure}[!tbp]
\begin{center}
% two columns
\includegraphics[clip,width=0.45\textwidth]{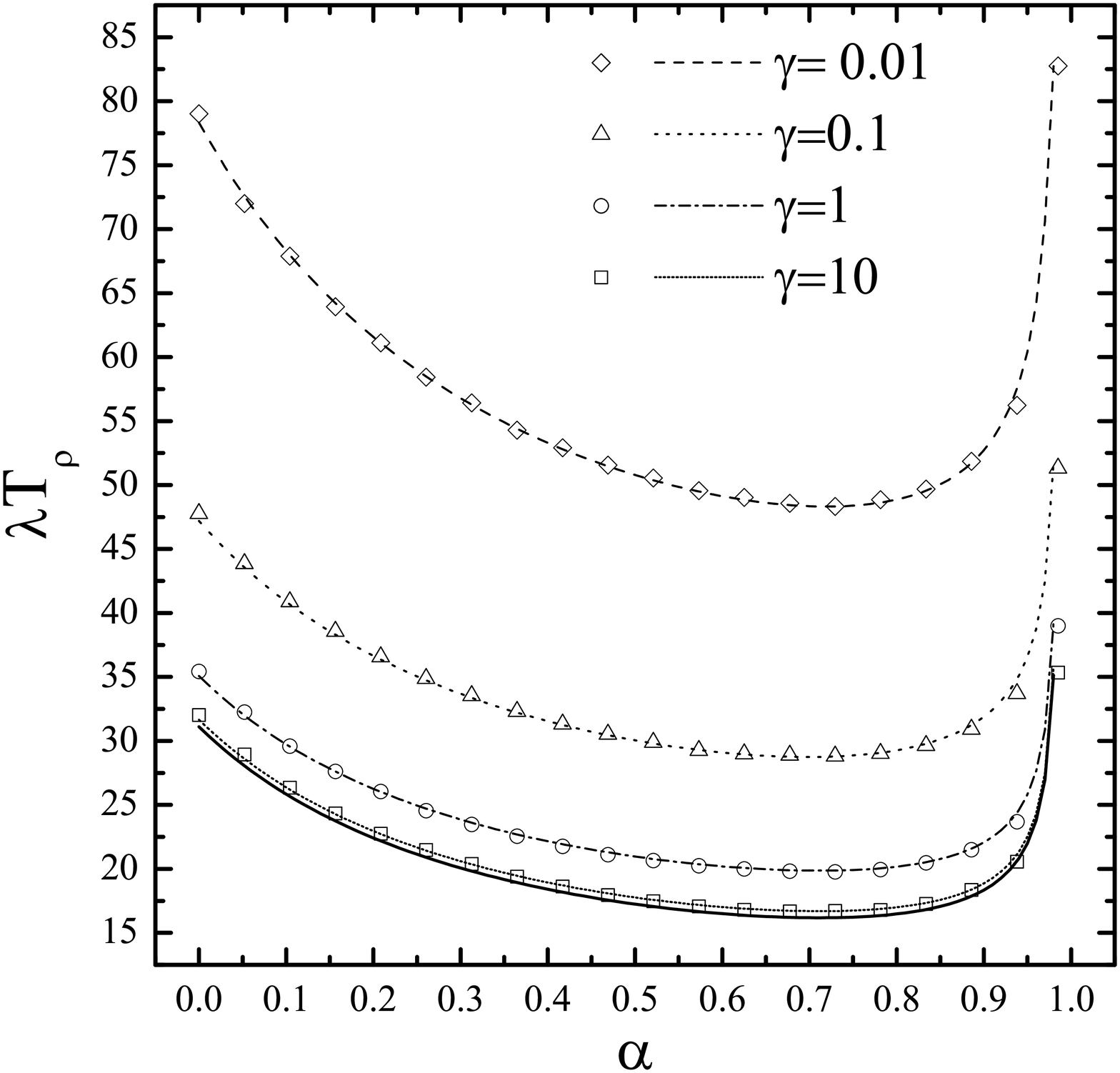}
% one column
%\includegraphics[clip,width=0.90\textwidth]{Fig_3.eps}
%
\caption{Analytical-Numerical calculations (lines) and Monte Carlo
simulations (symbols) for the Mean target lifetime (MTL) $T_{\rho}$,
for different target transition rates $(\gamma)$.
(I) (diamonds) $\gamma=0.01$, (II) (triangles) $\gamma=0.1$, (III)
(circles) $\gamma=1$ and (IV) (squares) $\gamma=10$. We have also
included for comparison the static trap case (thick solid line).}
\label{FigT1}
\end{center}
\end{figure}

\begin{figure}[!tbp]
\begin{center}
% two columns
\includegraphics[clip,width=0.45\textwidth]{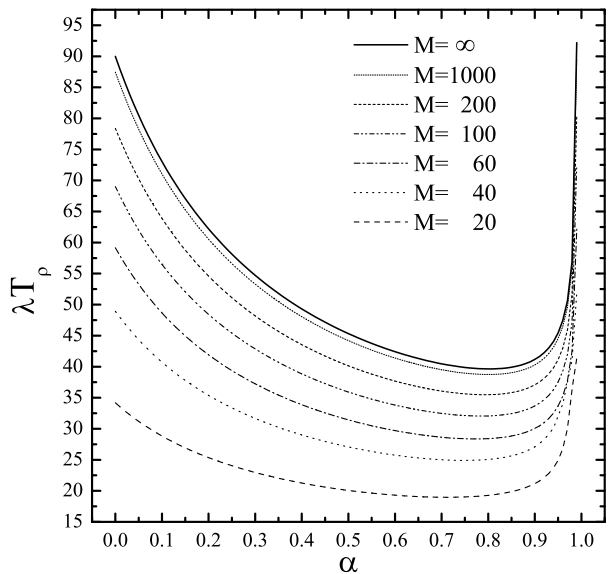}
% one column
%\includegraphics[clip,width=0.90\textwidth]{Fig_4.eps}
%
\caption{Analytical-Numerical calculations for the Mean Target
Lifetime (MTL) for a fixed target transition rate ($\gamma=1$)
as a function of the walker intermittency parameter $\alpha$
and for different sizes $M$ of the chain.
From bottom to top $M=20, 40, 60, 100, 200, 1000$ and $M=\infty$
(thick solid line).}\label{FigT2}
\end{center}
\end{figure}

The curves shown in Fig.~\ref{FigT1} correspond to the MTL,
in the finite case, as a function of the walker intermittency
parameter $\alpha$, for different target transition rates $\gamma$.
As it is clear from the figure, the results show the same trend
as the SP (Fig.~\ref{FigS1}) revealing also a remarkable rise in
MTL for low values of the target transition rates $\gamma$.
It could be inferred from the curves that with a modest transition
rate value ($\gamma= 0.1$) the target could almost double its
lifetime expectancy while a high ``activity'' of the target
($\gamma>1$) leads it to the static case.
Note that although MTL has less information than the SP, it shows
to be a simple and efficient tool for characterizing the proposed
search scheme.

\begin{figure}[!tbp]
\begin{center}
% two columns
\includegraphics[clip,width=0.45\textwidth]{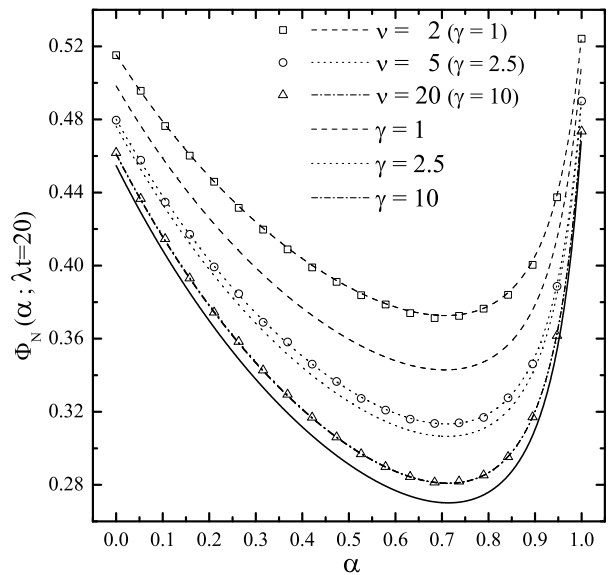}
% one column
%\includegraphics[clip,width=0.90\textwidth]{Fig_5.eps}
%
\caption{Analytical - Numerical calculations (lines) and Monte Carlo
simulations (symbols) for the SP, $\Phi(\alpha;t)$,
up to time $t=20$, in the high transition regime of the target.
Lines with symbols corresponds to the `imperfect' (high transition)
case and the same type of line (without symbols) for the dynamical
case. For instance (I) (squares) $\nu=2$, $\gamma=1$;
(II) (circles) $\nu=5$, $\gamma=2.5$ and (III) (squares) $\nu=20$,
$\gamma=10$. We have also included for comparison the static case
(thick solid line)}\label{FigS2}
\end{center}
\end{figure}

Figure \ref{FigT2} depicts the behavior of the MTL for a fixed target
transition rate ($\gamma=1$) as a function of the walker intermittency
parameter $\alpha$ and for different sizes ($M=20, 40, 60, 100, 200,
1000, \infty$) of the chain. In all cases was used a concentration
of searchers $\rho=0.1$. Notice how the finite chain (ring) approaches the infinite chain even for values
of $M$ not too large. As can be seen from the figure, the minimum in
MTL is maintained for all system sizes, which constitutes a robust
property of the intermittent search approach.
%Imperfect Trapping
In Fig.~\ref{FigS2} we consider the behavior of the SP,
$\Phi_N(\alpha;t)$, in the \emph{high} transition regime of the
dynamic target for a fixed evolution time $t=20$. In this limit,
the behavior of the SP approaches an imperfect trap
(see appendix \ref{AppendixB}), with
$\nu=(\gamma_1+\gamma_2)\gamma_2/\gamma_1$ being a ``measure of the
imperfection'' of the trapping process. When $\nu \rightarrow 0$
there is no trapping and if $\nu\rightarrow\infty$ perfect trapping
is achieved. Notice how the dynamical trapping resembles the imperfect
case even for values of $\gamma_i$ not too large.
It is worth remarking the excellent agreement between the
analytical--numerical results and the Monte Carlo simulations.

%%%%%%%%%%%%%%%%%%%%%%%%%%%%%%%%%%%%%%%%%%%%%%%%%%%%%%%%%%%%%%%%%%%%%%
\section{Conclusions}
%%%%%%%%%%%%%%%%%%%%%%%%%%%%%%%%%%%%%%%%%%%%%%%%%%%%%%%%%%%%%%%%%%%%%%

We have presented a simple model for the search kinetics of
a set of walkers performing intermittent motion in quest
for a dynamical target. The model is based only on RW, and our
results complement and extend previous related results given
in Ref.~\cite{050,055,060}. However, this model differs from those
mentioned.
In our previous work, the first searcher that finds the target,
captures it with probability one (we denominate that situation
the ``static trap'' in this work).
In the present work, an encounter walker-target does not necessarily
end in capture, but depends rather on the state of the dynamic target.

We have considered the target's survival probability (at a fixed time)
and the target's lifetime, and also studied the dependence
of these quantities on both the transition probability ($\gamma_i$)
between the states of the target and the parameter that characterizes
the walker's intermittency ($\alpha$).
Thus, we have established that the SP is a non-monotonic function
of $\alpha$ for a wide range of the transitions probabilities
$\gamma _1$ and $\gamma _2$, showing that intermittent strategies
still improve target detection when compared with the single-state
displacement. This confirms the utility of the intermittent search
approach~\cite{150} even in the case of a dynamical target.

We introduced the MTL and its connection with the SP was established.
As it was the case for SP, MTL was also a non-monotonic function of
$\alpha$ for several values of the transitions probabilities
$\gamma_i$, adequately depicting the improvement provided by
the intermittent search strategy.
Although MTL carries less information than the SP, it has shown to be
an efficient global optimizer for search strategies using intermittent motion.
In all cases the agreement between analytical-numerical results and
Monte Carlo simulations was quite good.

Thus, we have fulfilled our goal of presenting a simple model, based
only in diffusion, that captures in an unified framework the dynamical
behavior of the target and the intermittent search strategy
performed by the walkers.
The present scheme is both simple enough to be studied analytically
and rich enough to be able to mimic the influence of the
target's dynamics in the capture process and it shows that
intermittency is always favorable for optimizing the search.

The present model of intermittent search can be generalized in
several directions: higher dimensions, continuous systems,
non-Markovian target dynamics, etc. All of these aspects will
be the subject of future work.

%%%%%%%%%%%%%%%%%%%%%%%%%%%%%%%%%%%%%%%%%%%%%%%%%%%%%%%%%%%%%%%%%%%%%%
%%%%%%%%%%%%%%%%%%          Acknowledgments         %%%%%%%%%%%%%%%%%%
%%%%%%%%%%%%%%%%%%%%%%%%%%%%%%%%%%%%%%%%%%%%%%%%%%%%%%%%%%%%%%%%%%%%%%
\begin{acknowledgments}
The authors thank C.\ E.\ Budde Jr.\ for technical assistance
and wish to acknowledge support by CONICET and SeCyT
(Universidad Nacional de C\'ordoba), Argentina.
\end{acknowledgments}

%%%%%%%%%%%%%%%%%%%%%%%%%%%%%%%%%%%%%%%%%%%%%%%%%%%%%%%%%%%%%%%%%%%%%%
%%%%%%%%%%%%%%%%%%            Appendixes            %%%%%%%%%%%%%%%%%%
%%%%%%%%%%%%%%%%%%%%%%%%%%%%%%%%%%%%%%%%%%%%%%%%%%%%%%%%%%%%%%%%%%%%%%
\appendix

\section{}
\label{AppendixA}

Here, we describe some essential details of the calculations of
Section~(\ref{TrappingProcess}). We focus on the
$P_{i,i_0}(s, t|s_0,t=0)$, which are the building blocks for the
SP and MTL.
Given that we now apply our model only to chains, we drop out
the vector notation.

\subsection{Infinite Chain}
\label{InfiniteApp}

The solution of Eq.~(\ref{mastereq1}), with the
particularization of Eq.~(\ref{Aoperator}), for an infinite
(homogeneous) chain, can be given following the guidelines of
Refs.~\cite{110,155,160}.
The exploitation of the indicated formalism leads us to analytic
closed expressions for $P_{i,i_0}(s,t|s_0,t=0)$ in the
Fourier-Laplace space. However, as we are interested in
$P_{1,i_0}(s,t|s_0,t=0)$ (we take $i=1$ as the active status
of the target) we only show the solution for this case.
The relevant transformed results reads
\begin{eqnarray}
\nonumber\hat{P}_{1,1}(k,u)&=&\frac{1}{\Gamma}
\big(\gamma_2 \hat{P}^{0}(k,u) + \gamma_1 \hat{P}^{0}(k,u+\Gamma)\big)
\,, \\
\hat{P}_{1,2}(k,u)&=&\frac{1}{\Gamma}\big(\gamma_2\hat{P}^{0}(k,u) -
\gamma_2\hat{P}^{0}(k,u+\Gamma)\big) \,,
\end{eqnarray}
where $\Gamma=\gamma_1+\gamma_2$, and
$P^{0}(k,u) = 1/ (u-\mathbb{A}(k))$,
is the Fourier-Laplace transform of the conditional probability
$P^{0}(s,t|s_0,t=0)$, corresponding to the \emph{intermittent}
walker being at site $s$ at time $t$, given that it was at site
$s_0$ at $t=0$ (without trap) and
$\mathbb{A}(k)=\lambda[(1-\alpha)\cos{k}+\alpha\cos{2k}-1]$
is the Fourier transform of the evolution operator $\mathbb{A}$
given by Eq.~(\ref{Aoperator}).

We are interested in obtaining results for any \emph{initial} state
of the target/trap. Therefore is useful to evaluate the average
\begin{eqnarray}\label{P1Mar}
\nonumber \sum_{i_0}\hat{P}_{1,i_0}(k,u) \,\theta_{i_0}
= \hat{P}_{1,1}(k,u) \,\theta_1 + \hat{P}_{1,2}(k,u) \,\theta_2 \\
= \frac{1}{\Gamma}\left(\gamma_2\hat{P}^{0}(k,u)
+ (\gamma_1\theta_1-\gamma_2\theta_2)\hat{P}^{0}(k,u+\Gamma)\right),
\end{eqnarray}
where $\theta_1$ is the target probability of being initially
active, and $\theta_{2}$ of being inactive and satisfy
$\theta_1+\theta_2=1$. As usual, we choose for $\theta_i$
the equilibrium probabilities~\cite{090},
$\theta_1=\gamma_2(\gamma_1+\gamma_2)^{-1}$,
$\theta_2=\gamma_1(\gamma_1+\gamma_2)^{-1}$.
The Fourier inversion of Eq.~(\ref{P1Mar}) could be calculated
in an exact way resulting
\begin{equation}\label{PsMar}
\sum_{i_0}\hat{P}_{1,i_0}(s,u|s_0,0) \,\theta_{i_0}
= \ G\left(\frac{\eta_1^{|s-s_0|}}{\sqrt{x_1^2-1}}
+ \frac{\eta_2^{|s-s_0|}}{\sqrt{x_2^2-1}}\right) \,,
\end{equation}
where $\eta_{1}=x_1-\sqrt{x_1^2-1}$, $\eta_{2} = x_2+\sqrt{x_2^2-1}$,
$G={\gamma_2 / 2\lambda\Gamma\alpha(x_1-x_2)}  $
and
\begin{displaymath}
x_{1,2} = -\frac{1-\alpha}
{4\alpha}\pm\frac{1}{2}
\sqrt{\bigg(\frac{1-\alpha}{2\alpha}\bigg)^2
+2 \,\frac{u+\lambda(1+\alpha)}{\lambda\alpha}} .
\end{displaymath}
Averaging Eq.~(\ref{PsMar}) over the starting positions
(uniformly distributed) of the walker we arrive at
\begin{eqnarray}
\label{Paverage}
\sum_{\substack{i_0=1,2 \\s_0\neq0 }}
\hat{P}_{1,i_0}(s,u|s_0,0) \,\theta_{i_0}
\nonumber \\
= \frac{G}{\sqrt{x_1^2-1}}\frac{2\eta_1}{1-\eta_1}
+ \frac{G}{\sqrt{x_2^2-1}}\frac{2\eta_2}{1-\eta_2} \,.
\end{eqnarray}
Taking $s=0$, $s_0=0$, and considering Eqs.~(\ref{PsMar})
and~(\ref{Paverage}), we can write
\begin{equation}\label{Phi_inf}
\begin{array}{l}
\displaystyle
\sum_{s_0\neq0} \mathcal{L}
\left\{\left( 1 -  \Phi_{1}(0,t|s_0,0)  \right) \right\}
=\frac{1}{u} \sum_{\substack{i_0=1,2 \\s_0\neq0 }}
\hat{F}_{1,i_0}(0,u|s_0,0) \,\theta_{i_0}
\\
\displaystyle
=\frac{1}{u}\frac{1}{\hat{P}_{1,1}(0,u|0,0)}
\sum_{\substack{i_0=1,2 \\s_0\neq0 }}
\hat{P}_{1,i_0}(0,u|s_0,t=0) \,\theta_{i_0} \,.
\end{array}
\end{equation}
Eq.~(\ref{Phi_inf}) constitutes one of our main results and it allows
us derive SP from Eq.~(\ref{Phi_rho}) and MTL from Eq.~(\ref{MTLrho}),
after taking the inverse Laplace transform.
However, despite being able to obtain analytical results in Laplace
space for Eq.~(\ref{Phi_inf}), its length and complexity made the
analysis a difficult task. The analytical inversion of the Laplace
transform of the results seems to be beyond our possibilities,
so we have used a numerical procedure~\cite{145} for its calculation
in Sec.~\ref{results}.

\subsection{Ring of $M$ sites}
\label{RingApp}

For the finite case, we take the results from the previous
section, and obtain for the ring a solution in the form~\cite{170}
\begin{equation} \label{Pring1}
\hat{P}^M_{i,i_0}(s,u|s_0,t=0) =
\sum_{l=-\infty}^{\infty} \hat{P}_{i,i_0}(s+lM,u|s_0,t=0)\,.
\end{equation}
In order to evaluate the sum proposed in Eq.~(\ref{Pring1}),
we focus our attention in one term of Eq.~(\ref{PsMar}).
Given that for all $l\neq0$, $|l| \,M > (s-s_0)$
and if $l<0$  $|(s-s_0)+l \,M|=-(s-s_0) - l \,M$, we get
\begin{eqnarray}\label{ringT1}
\sum_{l=-\infty}^{\infty} \eta_{1}^{|s-s_0+lM|}
&=& \eta_{1}^{|s-s_0|}
+ \left( \eta{_1}^{|s-s_0|} + \eta_{1}^{-|s-s_0|} \right)
\sum_{l=1}^{\infty}\eta_{1}^{l M}
\nonumber \\
&=&\frac{1}{1-\eta_{1}^{M}}
\left(\eta_{1}^{|s-s_0|}+\eta_{1}^{M-|s-s_0|}\right) \,.
\end{eqnarray}
Working in a similar way with the other terms, we obtain the complete
solution of Eq.~(\ref{Pring1}), for the state of capture ($i=1$) as
\begin{eqnarray}\label{Pring2}
\sum_{i_0}\hat{P}^M_{1,i_0}(s,u|s_0,0) \,\theta_{i_0}
= \sum_{l=-\infty}^{\infty}\sum_{i_0}
\hat{P}_{1,i_0}(s+lM,u|s_0,0) \,\theta_{i_0}
\nonumber \\
=G \left(
\frac{\eta_{1}^{|s-s_0|}+\eta_{1}^{M-|s-s_0|}}
{\sqrt{x_1^2-1}(1-\eta_{1}^{M})} +
\frac{\eta_{2}^{|s-s_0|}+\eta_{2}^{M-|s-s_0|}}
{\sqrt{x_2^2-1}(1-\eta_{2}^{M})}
\right) \,.
\nonumber \\
\end{eqnarray}
\begin{widetext}
For a uniform distribution of walkers along the ring is useful
the expression
\begin{equation} \label{Paveragering}
\sum_{\substack{i_0=1,2 \\s_0\neq0 }}
\hat{P}^M_{1,i_0}(s,u|s_0,0) \theta_{i_0}=
G\left(
\frac{2}{\sqrt{x_1^2-1}}\frac{\eta_1-\eta_1^M}{(1-\eta_1^M)(1-\eta_1)}
+
\frac{2}{\sqrt{x_2^2-1}}\frac{\eta_2-\eta_2^M}{(1-\eta_2^M)(1-\eta_2)}
\right) \,.
\end{equation}
\end{widetext}
With Eq.~(\ref{Pring2}) valuated in $s=0$, $s_0=0$,
and taking into account (\ref{Paveragering}) we can write
\begin{eqnarray}\label{Phi_ring}
\sum_{s_0\neq0}\mathcal{L}
\left\{\left(1-\Phi_{1}(0,t|s_0,0)\right)\right\}
=\frac{1}{u}\sum_{\substack{i_0=1,2 \\s_0\neq0 }}
\hat{F}^{M}_{1,i_0}(0,u|s_0,0) \,\theta_{i_0}
\nonumber\\
=\nonumber\frac{1}{u}\frac{1}{\hat{P}^{M}_{1,1}(0,u|0,0)}
\sum_{\substack{i_0=1,2 \\s_0\neq0 }}
\hat{P}^{M}_{1,i_0}(0,u|s_0,t=0) \,\theta_{i_0} \,.
\\
\end{eqnarray}
From Eq.~(\ref{Phi_ring}), the SP (Eq.~(\ref{SurvivalN})),
$\Phi_{N}(t)$, and the MTL (Eq.~(\ref{MTLN})), $T_{N}$, are obtained.
However, as in the previous section, the size and complexity of
Eq.~(\ref{Phi_ring}) makes the inversion of the Laplace
transform beyond our possibilities, so we need to use a numerical
procedure~\cite{145} for obtaining the concrete results presented
in Sec.~\ref{results}.

\section{}
\label{AppendixB}

In this appendix we consider the {\em high transition regime
in dynamical trapping}, i.e., the behavior of the SP in the
limit $\Gamma \gg \lambda$. The calculation may be carried out
starting with the Laplace transform of Eq.~(\ref{PhiII}),
\begin{equation}\label{Phi1Lap}
\hat{\Phi}_{1}(0,u|s_0,0) = \frac{1}{u}
\left( 1-\sum_{i_0}F_{1,i_0}(0,u|s_0,0) \,\theta_{i_0} \right) \,.
\end{equation}
Let us consider the second term on the right hand side of
Eq.~(\ref{Phi1Lap})
\begin{eqnarray}
\sum_{i_0}\hat{F}_{1,i_0}(0,u|s_0,0)\theta_{i_0} =
\gamma_2 \,\hat{P}^{0}(0,u|s_0,0) \,/
\nonumber \\
\left( \gamma_2\hat{P}^{0}(0,u|0,0)
+\gamma_1\hat{P}^{0}(0,u+\Gamma|0,0) \right) \,.
\end{eqnarray}
In the considered limit ($\Gamma \gg \lambda$) ,
$\hat{P}^{0}(0,u+\Gamma|0,0)\sim1/\Gamma$~\cite{180}, then
\begin{eqnarray}\label{Apdimp}
\nonumber\sum_{i_0}\hat{F}_{1,i_0}(0,u|s_0,0) \,\theta_{i_0}
&\simeq&\frac{\gamma_2\hat{P}^{0}(0,u|s_0,0)}
{\gamma_2\hat{P}^{0}(0,u|0,0)+\gamma_1/\Gamma}\\
&\simeq&\frac{\nu\hat{P}^{0}(0,u|s_0,0)}
{1 + \nu\hat{P}^{0}(0,u|0,0)} \,,
\end{eqnarray}
where $\nu=\Gamma\gamma_2/\gamma_1$. Notice that Eq.~(\ref{Apdimp})
adequately provides the limits of perfect trapping
($\gamma_2/\gamma_1\gg1$, i.e., $\nu\rightarrow\infty$)
$\sum_{i_0}\hat{F}_{1,i_0}(0,u|s_0,0) \,\theta_{i_0}
=\hat{P}^{0}(0,u|s_0,0)/\hat{P}^{0}(0,u|0,0)$
and no target/trap present ($\gamma_2/\gamma_1\ll1$, i.e.,
$\nu\rightarrow0$)
$\sum_{i_0}\hat{F}_{1,i_0}(0,u|s_0,0) \,\theta_{i_0}=0$.
Using Eq.~(\ref{Apdimp}) and Eq.~(\ref{Phi1Lap}) we finally obtain
\begin{equation}\label{Phi1Imp}
\hat{\Phi}_{1}(0,u|s_0,0) \simeq
\frac{1}{u} \left( 1-\frac{\nu\hat{P}^{0}(0,u|s_0,0)}
{1 + \nu\hat{P}^{0}(0,u|0,0)} \right) \,.
\end{equation}
This result resembles the general case when detection of the
target upon encounter is less than certain, i.e.,
imperfect trapping~\cite{020}.
From Eq.~(\ref{Phi1Imp}) and using the same procedure from
Appendix~(\ref{RingApp}) the SP (Eq.~\ref{SurvivalN}),
$\Phi_{N}(t)$, and the MTL (Eq.~\ref{MTLN}),
$T_{N}$, could be evaluated for the present regime.

%%%%%%%%%%%%%%%%%%%%%%%%%%%%%%%%%%%%%%%%%%%%%%%%%%%%%%%%%%%%%%%%%%%%%%
%%%%%%%%%%%%%%%%%%             References           %%%%%%%%%%%%%%%%%%
%%%%%%%%%%%%%%%%%%%%%%%%%%%%%%%%%%%%%%%%%%%%%%%%%%%%%%%%%%%%%%%%%%%%%%
%\bibliography{Search}{}
%
%Merlin.mbs v4.21 2009-07-09.

%
%%%%%%%%%%%%%%%%%%%%%%%%%%%%%%%%%%%%%%%%%%%%%%%%%%%%%%%%%%%%%%%%%%%%%%
\end{document}